\documentclass[conference]{IEEEtran}

\usepackage[utf8]{inputenc}
\usepackage[T1]{fontenc}
\usepackage{lmodern}
\usepackage{url}
\makeatletter
\g@addto@macro{\UrlBreaks}{\do\/\do\-\do\_\do\.}
\makeatother
\usepackage{hyperref}
\hypersetup{
  pdftitle={Fifty Years of Specification Completeness: What Aviation Certification Tells AI Governance About Epoch Limits, Proof Surfaces, and the Structural Gap},
  pdfauthor={Christo Zietsman},
  pdfsubject={},
  pdfkeywords={AI governance, aviation certification, DO-178C, specification completeness, PromptQ, epoch limits, proof surfaces},
  colorlinks=true,
  linkcolor=black,
  citecolor=black,
  urlcolor=black
}
\usepackage{array}
\usepackage{booktabs}
\usepackage{microtype}
\usepackage{cite}

\title{Fifty Years of Specification Completeness: What Aviation Certification Tells AI Governance About Epoch Limits, Proof Surfaces, and the Structural Gap}
\author{Christo Zietsman \\ Nuphirho Research \\ \texttt{christo@nuphirho.dev}}
\date{2 June 2026}

\begin{document}
\maketitle

\begin{abstract}
Aviation software certification has operationalised three structural requirements for governed software systems since 1992: structured governance linkage between governing specifications and operational evidence, context-bounded validity that triggers revalidation when operational context changes, and an objective evidence architecture that defines what proof means and what makes it sufficient. These requirements appear in DO-178C and DO-330 and are enforced through FAA and EASA certification.

No existing framework requires these structural properties as intrinsic properties of individual AI governance documents. A system prompt, an AGENTS.md file, a governance policy, or a task envelope can be deployed without satisfying any of the three requirements aviation has enforced for three decades. Aviation is the most technically rigorous instance: its standard-setting bodies have acknowledged that their frameworks break down for AI systems, yet none requires these properties of individual governance documents.

Aviation's structural requirements break down at the system level because AI systems are non-deterministic, but remain transferable at the document level: the governance artifact is a static artifact whose structural properties can be evaluated independently of the stochastic system it governs.

The paper maps DO-178C's traceability architecture, DO-330's requalification triggers, and DO-178C's objective evidence requirements onto three structural findings: epoch limits on governance document validity, proof surfaces as the revalidation feedback mechanism, and the absence of structural completeness requirements in AI governance instruments. An empirical companion (arXiv:2604.21090) found that 37\% of AI governance documents fall below the structural quality threshold. PromptQ's seven-principle framework operationalises these requirements at the governance document layer.
\end{abstract}


\section{Introduction}

The problem of specifying software behaviour precisely enough to verify it is not new. Aviation software certification has been developing the engineering methodology for this problem since the publication of DO-178B in 1992 and its successor DO-178C in 2011. The core requirement is structured governance linkage: every software requirement must trace forward to a test objective and backward to a source system requirement. Extraneous code (code for which no requirement exists) is a certification finding. Undocumented behaviour (behaviour for which no test exists) is a certification finding. The specification is not a document that describes the software. It is a first-class artefact whose completeness the software must be provably derivable from.

AI governance documents have no equivalent requirement. A system prompt governing an AI agent can specify behaviour without defining what done looks like, how outputs will be verified, what the agent may refuse, or when the document's own authority expires. The document is deployed. The agent operates. No certification process checks whether the governing specification is complete.

This paper's primary conceptual move is to relocate the assurance question. Aviation's structural requirements are not transferable to AI systems at the system level: the aviation certification community has itself acknowledged in the CoDANN and NPA 2025-07 literature that DO-178C breaks down for non-deterministic AI because behaviour cannot be traced to requirements the way deterministic code can. The relocation this paper proposes is not to the AI system but to the AI governance artifact. The governance document (the system prompt, \texttt{AGENTS.md} file, governance policy, or task envelope) is a static authorship-time artifact. Its structural properties can be evaluated, required, and enforced independently of the stochastic system it governs. That decoupling is what makes the transfer viable.

The paper examines what aviation certification has required for thirty years, maps those requirements onto a framework of three research findings about AI governance document quality, and identifies the structural gap that separates the two traditions. The absence of structural completeness requirements for AI governance documents is indicated across nine regulated sectors and five language jurisdictions by a structured regulatory corpus audit of published standards and guidance instruments; this paper reports the aviation sector findings in detail.

\section{Background}
\label{sec:background}
\subsection{DO-178C and structured governance linkage}

DO-178C ~\cite{rtca2011do178c, faa2017} defines objectives, activities, and evidence requirements for five levels of software assurance (Design Assurance Levels DAL A through E). At DAL A (assurance level, applicable to software whose failure could cause a catastrophic aircraft accident) the standard requires full bidirectional traceability across four levels of specification hierarchy: system requirements, high-level software requirements, low-level software requirements, and source code.

The traceability requirement has two directions. Forward traceability: every system requirement must be traceable through software requirements to source code and to test cases. Backward traceability: every source code element must be traceable back to a software requirement, and every software requirement must be traceable back to a system requirement. Elements with no upstream requirement are classified as extraneous code and constitute a certification finding regardless of whether they function correctly.

This paper uses "structured governance linkage" rather than "bidirectional traceability" when describing the transfer of this principle to natural language governance documents. The distinction matters. DO-178C's bidirectional traceability requires formal syntax and deterministic compilation: source code can be objectively tested against requirements because code and requirements share a common formal substrate. Natural language governance documents do not share that substrate. The transfer this paper proposes is of the structural principle: that governance claims must be explicitly linked to evidence of their satisfaction, and that both gaps and extraneous content must be identifiable. This is not a transfer of the formal verification machinery. A natural language governance document achieves structured governance linkage when every claim is traceable to a verifiable evaluation mechanism and every evaluation mechanism traces back to an explicit governance claim.

A requirements set is complete under DO-178C when every downstream artefact is explainable by upstream intent and every upstream intent is evidenced downstream. This is an emergent property of the traceability network's structural closure, not a separately measurable attribute of individual requirements.

The quality criteria for requirements under DO-178C are explicit: requirements must be accurate, unambiguous, consistent, verifiable, traceable, appropriately detailed, and free of unintended functionality. These criteria map onto the seven structural principles of the PromptQ framework through functional correspondence, not formal equivalence:

\begin{table}[htbp]
\centering
\footnotesize
\caption{DO-178C requirements criteria mapped to PromptQ principles}
\label{tab:t1}
\begin{tabular}{p{4.1cm}p{3.5cm}}
\toprule
DO-178C requirements criterion & PromptQ principle \\
\midrule
Accurate (describes intended behaviour) & P1 Success Definition \\
Verifiable (provides a checkable criterion) & P2 Assessment Rubric \\
Free of unintended functionality & P3 Scope Boundary \\
Appropriately detailed on data sources & P4 Data Classification \\
Consistent with verification processes & P5 Quality Gate \\
Unambiguous (no multiple valid readings) & P6 Internal Consistency \\
Traceable (bounded to a context) & P7 Contextual Currency \\
\bottomrule
\end{tabular}
\end{table}

The mapping is not coincidental. Both frameworks are derived from the same underlying requirement: that a governance document must contain the components necessary for a conforming agent, whether a deterministic software system or an AI language model, to behave correctly, consistently, and within its currently valid scope.

\subsection{DO-330 and context-bounded validity}

DO-330~\cite{rtca2011do330} governs the qualification of software tools used in aviation certification. The DO-330 requalification trigger taxonomy provides a structural analogy, not a regulatory transplant, for governance document staleness. The analogy is between the conditions under which a certification tool loses its qualified status (and must be re-qualified before continuing to be used in the certification process) and the conditions under which a governance document loses its valid status (and must be revalidated before continuing to govern an AI system). Both involve artefacts whose continuing validity depends on the stability of the context for which they were originally validated. The governance document context is different from the tool qualification context, and the analogy is structural, not functional.

The requalification trigger taxonomy in DO-330 is the aviation framework's most precise implementation of what this paper terms an epoch limit: the condition under which a governance artefact's validity expires and revalidation is required. The taxonomy, documented in Pothon et al.~\cite{pothon2013}, distinguishes three cases:

Unchanged reuse: no requalification if TQL, lifecycle data, operational environment, tool operating requirements (TOR), and tool version all remain identical to the previously qualified configuration.

Operational environment change only: no requalification if the applicant demonstrates that the new environment is equivalent and the TOR remain applicable. The burden of proof is on the applicant; the default is invalidation.

Tool itself changed: requalification is required, scoped by an impact analysis identifying which aspects of the prior qualification remain valid and which must be renewed.

The critical boundary condition: the distinction between cases requiring requalification and cases not requiring it is not the presence or absence of change, but whether the change affects the relationship between tool outputs and certification objectives. A change that does not alter that relationship does not require requalification. A change that does alter it requires requalification regardless of its apparent magnitude.

The governance implication is direct: the AI governance document is the artefact whose validity must be context- bounded. When the deployment context changes in ways that affect the relationship between the document's governance claims and the agent's operational behaviour, the document must be revalidated. DO-330 places the burden of proof on the applicant to demonstrate contextual equivalence. No AI governance instrument currently imposes any equivalent burden.

\subsection{DO-178C objective evidence architecture}

DO-178C's treatment of evidence defines what it means for a claim to be supported rather than merely asserted. Evidence must:

\begin{itemize}
  \item support the certification objectives it is cited for, not merely be adjacent to them;
  \item survive independent review by someone other than the person who produced it;
  \item reconcile with all related lifecycle artefacts without contradiction;
  \item be produced through processes meeting the independence and rigour requirements appropriate to the DAL.
\end{itemize}

The adequacy-versus-presence distinction is explicit in the standard and in FAA Advisory Circular guidance: the existence of evidence does not make it sufficient. Evidence is sufficient when it demonstrates the property it is claimed to demonstrate under the conditions under which the claim applies.

DO-178C defines five evidence categories: review records, analysis records, test records, traceability data, and process evidence. Each category has defined objectives. Each objective is associated with independence requirements that scale with DAL level.

The programme's concept of the proof surface, the structured set of evidence mechanisms through which governance claims are verified, is a generalisation of this architecture to AI governance documents. Where DO-178C defines the proof surface for a software system's certification claims, the programme proposes that AI governance documents must define their own proof surface: the mechanisms through which their governance claims can be verified and the feedback loop through which deviations from those claims are detected and reported.

\section{The Gap}

\subsection{Where the aviation framework breaks down for AI}

The aviation certification community has itself identified where DO-178C fails for AI systems. CoDANN I~\cite{easa2020} and CoDANN II~\cite{easa2021} establish that ML systems challenge traditional certification because, as CoDANN II notes, "the link between requirements and implementation is statistical rather than deterministic." DO-178C's bidirectional traceability assumes deterministic behaviour: given the same input, a deterministic system produces the same output, and that output is either within or outside the specified behaviour. A neural network trained on data does not satisfy this assumption. Its behaviour is not a direct function of its requirements specification. There is no low-level design traceable from requirements to source code because the source code is the training process and the model weights, neither of which is specified by the requirements document. At inference time, sampling-based decoding means that even a fixed model presented with identical inputs does not guarantee identical outputs. The concept of unambiguous behaviour, central to both DO-178C's requirements criteria and any governance document's claim to be actionable, cannot be verified by the standard test-repeatability method when the system output is non-deterministic by design.

The absence is confirmed from within the most technically rigorous regulatory tradition. EUROCAE's ED-324 (in development, December 2026 target) is the first dedicated aviation AI governance standard. EASA's NPA 2025-07 proposes trustworthiness AMC/GM for AI in aviation. Neither specifies a structural completeness criterion for the governance document governing the AI system. The AIAA 2025-2511 paper ~\cite{lincoln2025} comes closest in peer-reviewed literature, arguing that AI requirements must be "clear, concise, and unambiguous." This addresses requirements quality, not governance document completeness.

\subsection{Where the gap appears}

The gap is not at the system level. The aviation framework has been actively developing AI-specific certification guidance for five years. The gap is at the document level: no aviation standard, advisory circular, or certification guidance specifies what must be present in the governance document (the system prompt, \texttt{AGENTS.md} file, or governance policy) that governs an AI system's behaviour before deployment.

The three structural requirements that DO-178C and DO-330 impose on software certification artefacts do not appear in any published AI governance instrument in aviation or in any other regulated sector. The programme has confirmed this across nine sectors and five language jurisdictions.

The aviation case is the sharpest illustration because aviation has the strongest tradition of formal specification completeness and the most developed methodology for implementing it. If the three requirements are absent from the most technically rigorous sector, their absence in less rigorous sectors follows with certainty.

\section{Three Findings Mapped}

\subsection{Epoch limits on governance document validity}

The first finding is that a governance document must carry its own validity conditions (the circumstances under which its governance claims remain applicable) and that revalidation is required when those conditions are violated. For continuously adapting systems, including those using retrieval-augmented generation or live data integration, evidence-based and event-based triggers are the appropriate epoch limit form: for example, "re-evaluate when the retrieval corpus is updated beyond a defined threshold." The framework requires declaration of the trigger; it does not prescribe its form.

DO-330's requalification trigger taxonomy is the most operationalised epoch limit specification in any regulated domain. It defines with precision what constitutes a change requiring revalidation (a change affecting the relationship between tool outputs and certification objectives), what constitutes a change not requiring it (a change demonstrably not affecting that relationship), and who bears the burden of proof (the applicant, not the certifier).

The programme's P7 principle (Contextual Currency) is the governance document analogue of this architecture. A P7-compliant governance document declares: here are the event-based conditions under which this document's governance claims expire and revalidation is required; here is the time-based interval after which revalidation is required regardless of events; and here is who is responsible for initiating revalidation.

No current AI governance instrument carries a P7-compliant staleness declaration. The default is implicit permanent validity. DO-330 makes the default the opposite: the default is invalidation, and the applicant must demonstrate continuing validity. The programme proposes the same inversion for AI governance documents.

\subsection{Proof surfaces as the feedback mechanism}

The second finding is that a governance document must define the mechanisms through which its governance claims are verified: what evidence is required, what makes that evidence sufficient, and how deviations are fed back into the governance document's revalidation cycle.

DO-178C's objective evidence architecture is the most developed proof surface specification in any regulated sector. Five evidence categories, defined objectives for each, independence requirements scaled to risk, adequacy- versus-presence distinction explicit in guidance.

The programme's P5 principle (Quality Gate), extended in v1.2 to require the document to name the evidence stream the human can independently access and to require active rather than passive human engagement, is the governance document analogue. The quality gate is where the proof surface is specified at authorship time.

The adequacy-versus-presence distinction has a direct governance document parallel: a quality gate that says "review outputs before accepting them" is present but not adequate. An adequate quality gate specifies what review means, what criteria outputs are reviewed against, and what action is required when criteria are not met.

\subsection{The structural gap confirmed}

The third finding is that no regulatory instrument in any confirmed jurisdiction requires AI governance documents to meet a structural completeness standard before deployment.

The aviation case provides the sharpest confirmation because it is the sector where the absence would be most surprising. DO-178C has required structural completeness of software requirements for thirty years. The standard- setting bodies adapting DO-178C for AI (EASA CoDANN, EUROCAE ED-324, NPA 2025-07) have acknowledged the framework's breakdown for AI but have not specified a structural completeness criterion for the AI governance document.

ED-324's acknowledged paradigm shift, from DO-178C's requirements completeness to ODD completeness, from code traceability to data/model lineage traceability, from functional determinism to statistical performance bounds, is a reframing at the system level. It does not address the governance document level. An AI system whose ODD is well-defined and whose data lineage is traceable can still be governed by a system prompt with no success definition, no quality gate, and no staleness declaration.

The gap is at the layer below the system and above the model: the natural language document that tells the system what to do.

The aviation finding is not isolated. A structured regulatory corpus audit of published governance instruments across eight further regulated sectors confirms the same absence. The three structural requirements (linkage, context-bounded validity, and proof surface specification) appear in none of the sector's primary governance instruments as requirements for AI governance documents. The primary instruments and the nature of the absence in each sector are as follows.

Financial services: SR 11-7 (Federal Reserve, 2011) is the most mature model governance framework in any sector. It requires model documentation, independent validation, and ongoing monitoring with defined performance thresholds for statistical models. It does not require the governance document for an LLM-based or agentic system to declare its own validity conditions or define a proof surface at authorship time. The monitoring requirement is a process obligation on the organisation, not a structural requirement on the document.

Healthcare: the FDA's Predetermined Change Control Plan (PCCP, 2021 and 2024 guidance) requires manufacturers to define in advance the types of changes an AI/ML algorithm may undergo and the performance specifications that trigger resubmission. This is the closest regulatory analogue to the epoch limit concept in any sector. The PCCP requirement applies to the change control plan as an organisational process document, not to the governance document specifying the AI system's operating scope before initial deployment.

Nuclear and critical infrastructure: IEC 61508 (functional safety for electrical, electronic, and programmable electronic systems) requires a safety case that must remain valid throughout the operational life of the system, with defined revalidation triggers when the operating context changes. This is the most operationally precise epoch limit requirement in any regulatory instrument. The IAEA safety case requirements impose the same obligation for nuclear systems. Neither instrument extends the safety case validity requirement to the governance documents for AI systems operating within or adjacent to safety-critical functions.

Legal and professional services: ABA Formal Opinion 512 (2023) requires lawyers to understand AI tool limitations and to supervise AI output. The UK Solicitors Regulation Authority guidance (2024) imposes equivalent obligations. Both instruments require the practitioner to exercise professional judgment as a compensating control for AI output. Neither specifies what structural properties the governance document for the AI tool must possess to make that judgment supportable.

Pharmaceutical and life sciences: ICH E9(R1) (2019) requires clinical trial sponsors to specify precisely what treatment effect they are estimating and under what conditions, before the trial begins. GxP computer system validation requirements (21 CFR Part 11, EU Annex 11) require documented evidence that validated systems do what their specifications say. Both instruments address the specification precision problem directly, in the context of clinical trial design and computer system validation respectively. Neither requires AI governance documents to satisfy structural completeness criteria before deployment.

Insurance and actuarial: the NAIC AI Systems Evaluation Tool (2023) and the UK PRA Supervisory Statement SS1/23 on model risk management in insurance mirror SR 11-7's model governance framework for the insurance sector. The NAIC instrument requires insurers to document governance, accountability, transparency, and consumer protection for AI systems. The structural completeness requirement, that the governance document itself must define what done looks like, how outputs are verified, and when the document expires, is not present in either instrument.

Public sector: the Five Eyes guidance "Careful Adoption of Agentic AI Services" ~\cite{cisa2026} recommends governance before deployment and human oversight throughout agentic AI operations in government contexts. The document explicitly acknowledges that "evaluation methods are not mature." That acknowledgment is the absence stated from inside the regulatory community: the instruments for confirming governance document completeness before deployment do not yet exist.

Management consulting: no sector-specific regulatory instrument governs AI governance documents produced by or for consulting firms. ISO/IEC 42001 is increasingly referenced as a baseline AI management system standard. It addresses organisational governance processes, not the structural properties of individual governance documents.

Multilingual confirmation: the absence is not limited to the Anglophone regulatory tradition. A structured audit of governance instruments in four additional language jurisdictions confirmed the same pattern. Japan's \textit{AI Guidelines for Business} Ver 1.2~\cite{meti2026}, France's CNIL AI Self-Assessment Guide~\cite{cnil2022}~\cite{cnil2022}, the South Korean AI Framework Act (enacted January 2025, in force January 2026), and China's SAC GB/T AI standards series none of which specify structural completeness requirements for AI governance documents as a precondition for deployment. This pattern is consistent across the regulatory traditions examined in this paper, with the aviation sector receiving the most detailed treatment.

\section{The Transferable Principles}

The aviation framework's failure for AI at the system level does not invalidate its principles at the governance document level. The three structural requirements are applicable to natural language governance documents for the same reason they are applicable to software requirements documents: both are specifications that a conforming agent must be derivable from, and both can fail in ways that are structural, identifiable, and consequential.

The aviation framework fails at the system level because AI systems are not deterministic derivations from their requirements. A neural network cannot be shown to trace to its requirements the way deterministic code can.

The aviation framework applies at the governance document level because natural language governance documents can be evaluated against structural completeness criteria just as requirements documents can. A system prompt that has no success definition fails the unambiguous criterion. A system prompt that has no quality gate fails the verifiable criterion. A system prompt that carries no staleness declaration fails the traceability criterion. These are document-level failures, not system-level failures, and they are assessable at authorship time.

A sceptic might object that this is no different from any informal specification, and that informal specifications have always been incomplete without any certification machinery to enforce them. The objection misidentifies the comparison. The relevant precedent is not source code but the natural language requirements documents that DO-178C has imposed structural criteria on for three decades. Those documents are also informal, also human-authored, and also subject to the same categories of failure: missing verifiability criteria, undefined scope, internal contradiction. DO-178C requires structural completeness of requirements documents precisely because their failures are detectable at authorship time and consequential at system level. The same logic applies to AI governance documents, with one additional force: in a deterministic software system, a requirements failure can in principle be detected during implementation when the code cannot be derived from the underspecified requirement. In a stochastic AI system, there is no implementation step that forces the failure to surface. The governance document is the last determinate artefact in the chain. If it is structurally incomplete, no subsequent process is guaranteed to catch it.

This is the programme's core claim: PromptQ moves aviation's structural requirements upstream, from the software artefacts that must satisfy them to the governance document that specifies what the software is supposed to do.

A further objection holds that document-level completeness cannot causally constrain a stochastic runtime. This is correct, and the framework does not claim otherwise. The claim is narrower and more defensible: a governance document that declares its own epoch limits creates an auditable trigger for human review when those limits are breached. A document that declares no triggers provides no such mechanism, regardless of what the runtime does. Unknown unknowns at authorship time cannot be specified; they are addressed not by completeness but by the re-evaluation mechanism the document must declare. When an unknown becomes known through operational experience, the specification is extended to incorporate the new learning. This is the Pyrrhonian inheritance: withhold assent on what cannot yet be known, but declare the conditions under which assent is warranted.

Where risk warrants it, the governance document can specify additional mechanisms: anomaly detection thresholds, runtime audit triggers, performance drift indicators. These extend the proof surface beyond authorship time without claiming to close it. No governance architecture, including DO-178C, provides absolute guarantees. Aviation's assurance model is not a claim of zero failures; it is a claim that failures are detectable, traceable, and correctable within a defined evidence architecture. The same standard applied at the governance document layer is not weaker than aviation practice. It is the same practice, applied one level upstream. Over time, as operational experience accumulates and unknown unknowns surface, the specification is extended, the proof surface grows, and the governance architecture approximates a robust and resilient solution without ever claiming to have reached one.

\section{ACAS Xu and the Assumption Epoch Limit}

The ACAS X collision avoidance system, specifically the ACAS Xu variant developed for unmanned aircraft integration into civilian airspace, provides a concrete example of aviation's handling of the epoch limit at the system level. ACAS Xu is a civilian airspace safety system, not a weapons platform; its assurance methodology is directly applicable to civil aviation governance contexts. It illustrates an additional precision that the programme's P7 principle should incorporate.

ACAS Xu uses a look-up table (LUT) as its specification: a formal verification that the neural network's output matches the LUT across the entire input space. This addresses completeness by verifying the specification- approximation gap is zero. But the assurance is bounded by the validity of the LUT's underlying assumptions: perfect sensors, bounded manoeuvre models, cooperative intruder assumptions.

The assumption-centric assurance approach treats the assumptions register (the explicit list of conditions under which the assurance holds) as the specification. When those assumptions no longer hold, the epoch expires. When the sensors are imperfect, or the intruder is non-cooperative, or the manoeuvre bounds are exceeded, the assurance case is invalid regardless of whether the neural network still matches the LUT.

This identifies a third type of epoch limit beyond the time-based and event-based triggers in P7's current definition: an assumption epoch limit. A governance document becomes invalid when the foundational assumptions under which it was written no longer hold, even when no discrete triggering event has occurred and no calendar interval has elapsed. A P7-complete governance document should declare its foundational assumptions alongside its staleness declaration.

The assumption-based trigger is a proposed research direction. It is less mature than the time-based and event-based triggers, which have clear operational precedents in aviation certification practice. Formalising the assumption epoch limit requires a taxonomy of governance document assumption types, monitoring mechanisms, and invalidation thresholds. The ACAS Xu case provides the assurance-structure motivation; the operational specification is a direction for further work.

The epoch limit concept is derived from the structural principle that assumptions have validity boundaries, not from ACAS Xu's specific algorithmic mechanism. ACAS Xu illustrates the principle in a deterministic system; the governance document application extends it to the authorship-time artefact layer where the analogous mechanism is human review triggered by declared conditions rather than algorithmic state transition.

\section{Implications for AI Governance Practice}

The aviation case provides three actionable implications for AI governance practice.

First: governance documents should be treated as certification artefacts, not operational instructions. DO-178C treats the requirements document as a first-class artefact with its own quality criteria, completeness requirements, and traceability obligations. AI governance practitioners should apply the same standard to the documents that govern AI system behaviour.

Second: the burden of proof for continuing validity should be on the deployer, not the certifier. DO-330 makes invalidation the default; demonstrating continuing validity requires active argument. AI governance practice has the opposite default: deployment is assumed valid until something goes wrong. The aviation inversion (assume invalid, demonstrate valid) is the appropriate default for high-stakes AI systems.

Third: evidence architecture should be specified at authorship time, not assembled retrospectively at audit. DO-178C requires that evidence categories and objectives be defined as part of the development process. The evidence is produced as the system is built. AI governance practice typically assembles evidence after deployment when compliance is questioned. Defining the proof surface at authorship time: what evidence will demonstrate compliance, what makes that evidence sufficient, how deviations will be fed back. This is the aviation model transferred to AI governance.

The assurance case tradition provides established notation for this ~\cite{bloomfield2010}. Goal Structuring Notation ~\cite{kelly2004} is the standard method for expressing structured arguments that a system is safe with respect to a given set of requirements, linking claims through explicit inference rules to a defined evidence base. An AI governance document that declares its proof surface is making the same structural move: it specifies the claims, the inference structure connecting those claims to evidence, and the evidence required to make each claim hold. The difference is that GSN was developed for safety-critical systems engineering. The concept of a proof surface in this paper is closely related to GSN's evidence layer; the novel contribution is the requirement that individual AI governance documents declare their own proof surface at authorship time, rather than having it constructed retrospectively as part of a system-level safety argument. The proposal is to require this structure in the natural language governance documents that precede any engineering, not to reproduce GSN as a notation system.

\section{Related Work}

The gap between DO-178C's requirements and AI system assurance has been studied in the CoDANN I~\cite{easa2020} and CoDANN II~\cite{easa2021} literature. These documents establish the breakdown of the traditional certification model for learning systems. The ED-324 standard in development (EUROCAE, target December 2026) is the primary ongoing standardisation response.

The ACAS Xu assumption-centric assurance approach is documented in Damour et al.~\cite{damour2021}, Gabreau et al. (2022, ERTS), and Gabreau, Teuli\`{e}res et al.~\cite{gabreau2024}. These papers represent the most developed technical work on assumption-bounded AI certification in aviation.

The programme's companion paper (arXiv:2604.21090~\cite{zietsman2026b}) established that 37\% of AI governance documents fall below a structural quality threshold measured against the five-principle diagnostic model. Applying a seven-principle extension of this framework, adding P6 (Internal Consistency) and P7 (Contextual Currency), to the same 34-file corpus (the governance-prompts-v1 empirical corpus) yields the following results.

The table below shows mean scores by principle across the 34-file corpus under the seven-principle extension. 94\% of files (32/34) score below the 3.5/7 threshold.

\begin{table}[htbp]
\centering
\footnotesize
\caption{Seven-principle mean scores, 34-file corpus (five raters)}
\label{tab:t2}
\begin{tabular}{p{4.1cm}p{3.5cm}}
\toprule
Principle & Mean score \\
\midrule
P1 Success Definition & 0.44 \\
P2 Assessment Rubric & 0.41 \\
P3 Scope Boundary & 0.46 \\
P4 Data Classification & 0.24 \\
P5 Quality Gate & 0.47 \\
P6 Internal Consistency & 0.60 \\
P7 Contextual Currency & 0.00 \\
\bottomrule
\end{tabular}
\end{table}

The P7 finding is the most significant result in the programme. Every governance document in the 34-file corpus scores zero on Contextual Currency. Not a single document declares any staleness trigger (event-based, time-based, or evidence-based). The score is 0.00 with zero variance, confirmed across five independent raters from two model families. The corpus represents real-world practice: no practitioner in the evaluated corpus has specified when their governance document expires. This is the empirical basis for the theoretical claim this paper develops: the absence of epoch limit declarations is not an edge case in production AI governance; it is the norm.

The increase from 37\% (five-principle) to 94\% (32/34) (seven-principle) is structural. P7 contributes 0 to every file; P6 adds up to 1 point of exposure; the higher threshold (3.5/7 versus 2.5/5) reflects the extended ceiling. The five-principle rate remains valid as a lower bound for the P1-P5 failure profile. The seven-principle rate captures the full structural gap including the epoch limit dimension this paper addresses.

Concurrent work on evidentiary AI governance frameworks includes OpenKedge (He and Yu, arXiv:2604.08601, 2026) and Koch and Wellbrock~\cite{koch2026}. These papers address the runtime evidence layer; this paper addresses the authorship-time specification layer.

\section{Conclusion}

Aviation software certification has required structured governance linkage, context-bounded validity, and an objective evidence architecture for three decades. These requirements have made aviation software among the most reliable in any safety-critical domain.

No equivalent requirements apply to the governance documents that specify AI system behaviour. The aviation certification community has acknowledged that its existing frameworks break down for AI systems at the system level, and the standards being developed to address that breakdown have not yet specified structural completeness requirements for AI governance documents. This absence is indicated across nine regulated sectors by a structured regulatory corpus audit; the aviation sector provides the sharpest instance because it has the most technically rigorous specification tradition.

The paper's primary claim is that aviation's structural requirements are transferable to AI governance documents at the document level, even though they cannot be transferred to AI systems at the system level. A governance document is a static authorship-time artifact. Its structural properties can be evaluated and required independently of the stochastic system it governs. The seven-principle PromptQ framework is the methodological operationalisation of that transfer.

A circularity limitation applies to the empirical results. The PromptQ instrument was designed by the author; using it to score a corpus and find a gap does not independently validate the instrument. Three mitigants are noted. First, each of the seven principles is grounded in an established external tradition: safety engineering (P1, P3, P5), measurement theory (P2, P4), and organisational theory (P5, P6, P7), providing independent theoretical justification for why each property is required. Second, the P7 finding (zero documents declare any staleness trigger across five independent models from two model families) is consistent regardless of the scorer's leniency gradient, and holds under per-principle scoring modes that suppress holistic evaluation bias. Third, the three structural requirements identified in Section II are derived from aviation certification practice, not from PromptQ. PromptQ operationalises a pre-existing standard, it does not define it. Independent empirical validation of PromptQ against established requirements engineering metrics is potential future work.

The absence of equivalent requirements in AI governance reflects a policy choice not to require even the structurally feasible completeness properties that aviation's tradition identifies. Whether governance documents can achieve full DO-178C-equivalent traceability at scale remains an open research question requiring formalism not yet developed for natural language artifacts. The paper argues only that the structural principle is transferable and the absence is not inevitable.

Risk-proportionate application is the appropriate target. DO-178C's Design Assurance Level structure scales evidence requirements with hazard severity: DAL A for catastrophic failure, DAL E for no safety effect. Governance document completeness requirements should scale similarly. The PromptQ framework's graduated score (0-7) provides the natural substrate for risk-tiered minimum thresholds. The table below proposes provisional thresholds; precise calibration requires empirical grounding and is a primary direction for further work.

\begin{table}[htbp]
\centering
\footnotesize
\caption{Provisional risk-proportionate PromptQ minimum score thresholds}
\label{tab:t3}
\begin{tabular}{p{1.9cm}p{2.4cm}p{3.1cm}}
\toprule
Risk tier & Indicative minimum PromptQ score & Notes \\
\midrule
Prohibited (EU AI Act) & Not applicable & Document completeness cannot substitute for prohibition \\
High risk (EU AI Act Annex III; NIST critical) & 6/7; P1, P3, P5, P7 mandatory & All staleness triggers required; scope boundary mandatory \\
Limited risk (EU AI Act transparency obligations) & 4/7; P1, P3 mandatory & Success definition and scope boundary are the minimum viable properties \\
Minimal risk & 3.5/7; no mandatory principles & Programme corpus threshold \\
\bottomrule
\end{tabular}
\end{table}

The empirical corpus shows that 94\% of current practitioner documents fall below 3.5/7; a high-risk threshold of 6/7 would be failed by every document in the corpus. Whether this reflects a calibration problem or a real governance gap is itself a research question that requires outcome data linking document scores to governance failures.

\end{document}